# AI-Research-Lens augmented bibliometric framework: A paradigm shift with agentic AI for dynamic, snippet-based research analysis

Adela BÂRA, Simona-Vasilica OPREA*
*Department of Economic Informatics and Cybernetics, Bucharest University of Economic Studies, Bucharest, Romania*
*Corresponding author. E-mail: simona.oprea@csie.ase.ro

**Abstract:** Our paper introduces a generative, multi-agent AI framework designed to overcome the limited flexibility and technical barriers of current bibliometric tools. The objective is to enable researchers to perform fully dynamic, code-based scientometric analysis using natural-language (NL) instructions, eliminating the need for specialized programming skills while expanding analytical depth to hypotheses discovery. Methodologically, the system integrates four coordinated AI agents: a custom analytics generator, a full-paper retriever, including a Retrieval-Augmented Generation (RAG)-based researcher assistant and an automated report generator. User queries are translated into executable Python scripts, run within a sandbox ensuring safety, reproducibility and auditability. The framework supports automated data cleaning, construction of co-authorship and citation networks, temporal analyses, topic modeling, embedding-based clustering and synthesis of research gaps. The novelty lies in unifying NL-to-code scientometrics, full-paper retrieval, agentic exploration and hypotheses generator, dynamic metric creation in a single adaptive environment, capabilities that are not typically integrated within existing bibliometric platforms such as VOSviewer, Bibliometrix or SciMAT. Unlike static GUI-based workflows, the proposed framework supports iterative "what-if" analysis, hybrid indicators and user-driven pipeline modification. Results demonstrate that the framework generates valid analysis scripts, retrieves and synthesizes full papers, identifies research themes and produces reproducible scientometric outputs.

## 1. Introduction

### 1.1 General context, importance of the proposed framework and current gaps

Despite the proliferation of advanced scientometric techniques, embedding-based clustering, temporal network analysis, full-text mining, many researchers remain limited to predefined functions available in VOSviewer, Bibliometrix or SciMAT [1]. These tools, while powerful, lack flexibility: users cannot easily define new metrics, combine heterogeneous data sources or explore iterative "what-if" analyses without writing custom code. LLMs lower this barrier by enabling on-the-fly generation of analytical scripts, automated data cleaning and interactive exploratory workflows. With the growing availability of large bibliometric datasets, an intelligent system capable of transforming NL into executable scientometric pipelines becomes timely and necessary [2], [3], [4].

Despite the maturity of bibliometric methodologies, current tools still exhibit significant limitations that restrict their analytical power and accessibility. Most platforms offer static functionality, constraining researchers to predefined metrics and fixed analytical workflows that cannot accommodate novel, hybrid, or domain-specific indicators. Interactivity is also limited: existing software generally lacks the ability to adapt pipelines based on iterative user refinements, exploratory prompting or dynamic analytical intent [5], [6]. Moreover, traditional approaches depend heavily on metadata while providing poor integration with full-text content, thereby overlooking the semantic richness present in abstracts, methodologies, findings and discussions. Further, advanced analyses, such as temporal embeddings, graph-based algorithms and topic modeling, require substantial programming expertise, creating a high technical barrier that excludes many researchers [7], [6]. At the same time, current tools offer little personalization, making it difficult to tailor analyses to specific research questions, fields or evolving datasets. Reproducibility remains another critical bottleneck, as most bibliometric pipelines lack transparent code execution histories, automated notebook generation or systematic tracking of analysis steps [8].

### 1.2 Contributions and novelty

To address these shortcomings, our work proposes a generative and agentic AI framework (named AI-Research-Lens) for dynamic, code-based bibliometric and research analysis, introducing several novel contributions. Four AI agents are designed: (1) Custom Analytics Generator (CAG), (2) Paper Retriever (PR), (3) Research Assistant (RA) based on RAG technology, (4) Report Generator (RG). The emergence of large language models (LLMs) offers a transformative opportunity to reimagine this landscape by enabling fully dynamic, interactive and scalable bibliometric intelligence [9], [10]. Our research proposes an AI-augmented, agentic and generative AI-based framework that allows users to perform dynamic, snippet-based, fully code-driven bibliometric analyses through Natural Language (NL) queries and discover new research hypotheses based on identified gaps. The objective is to design an end-to-end system in which researchers articulate analytical goals in NL, a LLM generates valid Python code tailored to the request, a secure execution environment (sandbox) runs the code and the system returns figures, networks, topic models and other statistics, then identify research gaps and propose new hypotheses allowing users to retrieve full papers to further investigate these hypotheses. The framework aims to democratize scientometric analysis by eliminating the need for advanced programming, while simultaneously expanding analytical depth through generative AI capabilities.

AI-Research-Lens acts as a lens over complex scientific corpora, enabling users to explore, analyze and synthesize knowledge from large collections of documents. The framework enables NL-to-code Scientometrics, where a multi-agent LLM pipeline translates user instructions directly into executable Python scripts, allowing real-time generation of custom analyses without requiring programming skills. To ensure transparency, every analytical session is coupled with full reproducibility, automatically generating a complete Python notebook that captures code, reasoning steps and results.

While LLM-based code generation and RAG mechanisms have been previously explored in isolation and in developer-oriented tools such as GitHub Copilot and Cursor, our contribution lies in their unified integration into an agentic, end-to-end bibliometric research system. The proposed framework illustrates how generative AI and agent-based architectures support more flexible and interactive scientometric workflows compared with traditional static pipelines. It autonomously generates, executes, validates and documents analytical pipelines, integrates full-text semantic reasoning and generates hypotheses based on the analyses conducted for non-programmer researchers.

Beyond metadata-based workflows, the system integrates a *full-paper* retriever of open access publications to build a knowledge base corpus that supports further exploration and investigation of existing literature to document the newly proposed hypotheses. A RAG agent receives user prompts and retrieves data from the corpus, generating responses using LLMs. It enables advanced text interpretation through agentic AI components capable of identifying frontier themes, interpreting methodological workflows and models/algorithms implementation, synthesizing research gaps and extracting useful insights. Methodologically, the proposed framework establishes the first agentic architecture for bibliometrics and research analytics that integrates sandboxed execution, automatic code generation for reproducibility, semantic enrichment via embeddings, full-text-aware topic interpretation and intelligent synthesis of research gaps.

## 2. Literature review

### 2.1 Existing bibliometric platforms

Comparative reviews highlight tools such as VOSviewer [11], CiteSpace [12], Bibliometrix [13], Biblioshiny, SciMAT [14], Sci2, CitNetExplorer and BibExcel as the dominant platforms for constructing and visualizing citation, co-authorship and keyword co-occurrence networks, each with distinct strengths in preprocessing, visualization and export capabilities. Aria and Cuccurullo's bibliometrix package extended these capabilities within R, enabling end-to-end science mapping, but still presupposing familiarity with scripting and statistical environments [13]. SciMAT similarly offers advanced longitudinal mapping and thematic evolution analysis through a graphical interface, yet its workflows are largely pre-defined and difficult to reconfigure for bespoke indicators or hybrid metrics [14].

Recent work has also examined how these tools are actually used in published studies, documenting typical workflows and recurrent methodological weaknesses, such as insufficient reporting of parameter settings and data-cleaning procedures, which ultimately limit reproducibility [15]. At the same time, a growing body of domain-specific bibliometric studies on AI and machine learning (ML) [16], NL processing and LLMs in particular, demonstrates the versatility of classical scientometric techniques, but also reinforces their reliance on relatively rigid, script- or GUI-based framework reliably generates [17], emphasizing methodological fragmentation, limited interoperability and challenges in reproducibility [18], [19]. In parallel, significant advances in citation indicators, name disambiguation [20] and network-based scientometric metrics continue to refine the analytical toolkit available to researchers [21].

A comprehensive review by Moral-Muñoz et al. shows that most bibliometric tools fall into three categories: general performance analysis, science mapping tools and libraries, and stresses that current ecosystems remain fragmented, with limited interoperability and substantial technical barriers [22]. More recent work using VOSviewer as a "diagnostic lens" for exploratory bibliometrics confirms that meaningful analysis often still depends on expert parameter tuning and manual interpretation rather than adaptive, interactive guidance [23].

**2.2 Bibliometric embedding ML and AI**

Studies on "data-driven research impact science" show how clustering, topic models and advanced embeddings can be combined with traditional citation indicators to obtain richer, multi-dimensional views of impact [24], [25], [26]. Other work proposes AI-augmented bibliometric analysis, arguing that algorithmic support is increasingly necessary to handle the scale and complexity of modern scholarly corpora, while still typically treating AI as a back-end component embedded in fixed tools rather than as an interactive, programmable layer [27]. As scientometrics community has started to incorporate more ML and AI within traditional, researcher-coded pipelines, Hain et al. review how ML and AI have been used for science, technology and innovation mapping and forecasting, emphasizing their potential for topic evolution modeling, trend detection and early-warning indicators, yet these approaches remain technically demanding and largely inaccessible to non-experts [28]. Further, Oyewola and Dada rely on bibliometrix and VOSviewer to produce maps and clusters, but the workflow is still built from a set of fixed functions with limited room for dynamic metric design [29]. Additionally, Glenisson et al. showed early on that combining full-text and bibliometric information can yield richer term maps and more nuanced field delineation [30]. More recently, Balz conducted a full-text scientometric analysis of all papers in *Remote Sensing* (2009-2021), demonstrating that large-scale text mining enables detailed thematic profiling that would be difficult to obtain from titles and abstracts alone [31].

Velásquez's TechMiner provides a Python-based environment for cleaning Scopus data, disambiguating authors and running advanced analyses, such as keyword clustering and citation analysis, explicitly positioning itself as a flexible alternative to "black-box" GUI tools [32]. Spinellis' Alexandria3k goes further by offering fast relational access to large, openly available publication datasets and promoting fully reproducible scientometric workflows via public code and replication packages [33]. Kluyver et al. present Jupyter Notebooks as a publishing format that tightly couples code, narrative and results, while Ziemann et al. formalize the "five pillars" of reproducible computational research [34].

**2.3 Generative AI and NLIs**

Systematic reviews discuss the use of generative AI in higher education and computer-science education, the integration of generative tools into academic research workflows and peer review [35] and the opportunities and risks of generative AI for scientific writing and literature reviews [36]. Empirical evaluations have assessed ChatGPT and related models as real-time literature search assistants, examined the accuracy and reliability of AI-generated citations and references [37] and mapped the broader implications of generative AI for scientometrics as a field [38]. Integrated scholarly-data platforms such as MetaInfoSci point toward web-based environments that combine automated workflows, network analytics and interactive dashboards to support large-scale, reproducible bibliometric studies, while still relying on pre-defined modules rather than dynamically generated analysis code [39], [40], [41].

Srinivasan and Stasko show that NLIs for data analysis with visualization can help users express analytical intent without grappling with complex widgets or query languages, though current systems still handle only relatively constrained classes of questions [42], [43]. Voigt et al. survey NL interaction in visualization more broadly, cataloguing how NL can be used not only for querying but also for annotations, recommendations, explanations and analytical conversation [44]. Shen et al. provide a comprehensive survey of NLIs for data visualization, proposing a multi-stage pipeline from query interpretation to visual mapping and interaction management and identifying the need for systems that can support rich, multi-turn analytical dialogues [45].

The progress of LLMs, especially in code generation, creates new opportunities to link these strands. Yin et al. introduce ARCADE, a benchmark for NL-to-code generation in interactive data science notebooks (pandas) and show that specialized code models can translate multi-turn NL intents into executable analysis scripts within notebooks [46]. Xu et al. study in-IDE code generation from NL and find that ML-based assistants can help developers, but also highlight persistent issues with correctness, usability and user trust, underscoring the need for validation and safe execution environments [47].

Building on these foundations, Jansen et al. systematically evaluate how well general-purpose LLMs automate data analysis workflows, showing that while simple tasks often yield executable code, task complexity sharply reduces reliability and reinforces the importance of guardrails, prompt engineering and human oversight [48]. Nejjar et al. focus specifically on scientific use cases for LLMs, documenting how researchers already employ LLM-based tools for code generation and data analysis, but also reporting substantial variability in output quality and integrity across tools and tasks [49]. A brief comparison of the previous works is provided in Supplementary material, Table S1.

Despite substantial advancements, the literature identifies consistent limitations: reliance on static analytical modules, lack of interactivity and adaptive workflows, barriers for non-programmers performing advanced analyses, limited integration of full-text semantic features and weak reproducibility. To date, however, no unified framework integrates: (1) custom analytics based on code generation, including sandboxed execution of dynamically generated scientometric pipelines, (2) research-driven agent that identify gaps, trends and proposes new research hypotheses, (3) full-paper retriever using OpenAlex[1] API, (4) report generator based on the analytics. These gaps motivate the AI-augmented, multi-agent framework proposed in our paper, which explicitly positions generative models as the core mechanism for dynamic code synthesis, snippet-based bibliometric querying and end-to-end reproducible scientometric workflows.

## 3. Methodology

The proposed framework is conceived as a hybrid, multi-layered architecture that unifies classical bibliometric analytics with an agentic, AI-driven discovery pipeline. Unlike traditional bibliometric systems that provide descriptive and network-based insights, our approach introduces a second, reasoning-oriented layer that transforms analytical outputs into actionable research exploration. The framework consists of two layers: the Analytics Layer performs the full spectrum of established scientometric processing, while the Agentic AI Layer builds upon these results to autonomously identify research gaps, infer emerging trends and generate structured research hypotheses.

At the core of this integration lies a sequential knowledge amplification process. The structured bibliometric data are processed through the Analytics Layer, which produces descriptive indicators, network representations and topic structures. These outputs are not treated as final results but as intermediate, machine-interpretable knowledge artifacts. The Agentic AI Layer consumes this enriched analytical state and performs higher-level reasoning tasks, including gap detection, thematic synthesis and hypothesis generation. In this way, the system shifts from static analysis toward a dynamic, hypothesis-driven research paradigm.

A key innovation of the framework is the transformation of generated research hypotheses into executable discovery actions. The research hypotheses are converted into structured search queries, which are executed via the OpenAlex API to retrieve relevant, open-access scientific publications. This creates a closed-loop interaction between analysis and discovery, where insights derived from bibliometric data directly drive the expansion of the underlying knowledge base. The retrieved documents are subsequently processed, parsed and embedded into a vector representation space. To support scalable semantic retrieval and contextual reasoning, all newly acquired documents are ingested into a FAISS-based vector index, forming a continuously evolving knowledge repository. The RAG mechanism allows the system to ground its responses in both the original bibliometric dataset and the dynamically expanded corpus of full-text publications, thereby enhancing the depth, reliability and interpretability of generated insights.

[1] https://openalex.org/

Through this iterative pipeline, the framework establishes a feedback loop between analytics, hypothesis generation, literature retrieval and semantic reasoning. Each cycle enriches the system's knowledge base and refines subsequent analyses, effectively enabling continuous, data-driven exploration of scientific domains (as in Figure 1).

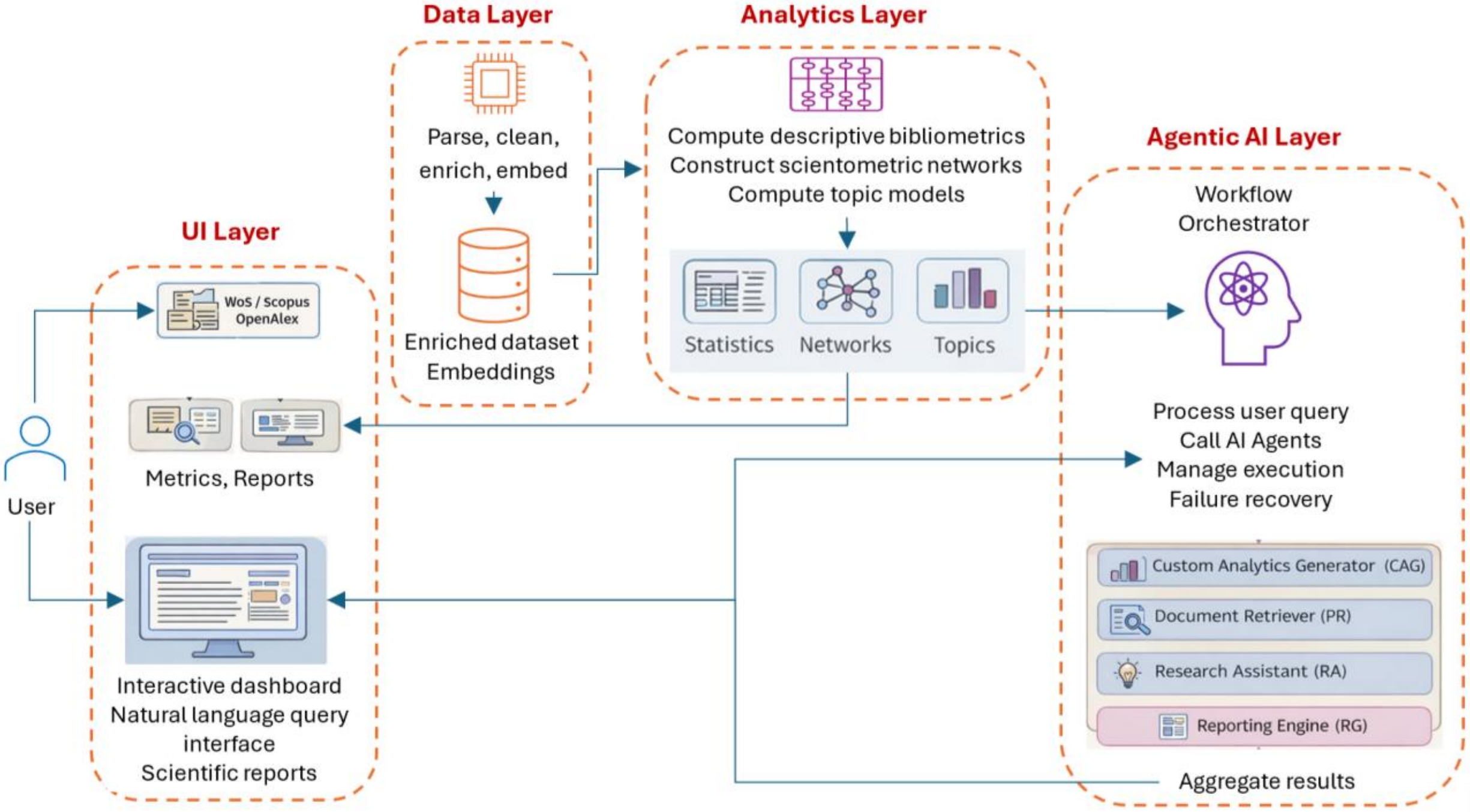

Figure 1. The architecture of the proposed framework

### 3.1 Data Layer

This layer handles structured bibliographic data from records obtained primarily from WoS Core Collection, Scopus and OpenAlex. Users upload bibliometric datasets which are parsed through a standardized import module. The preprocessing pipeline applies the following steps: i) normalizes author names, ORCID identifiers, institutional names and country fields; ii) uses DOI-based matching combined with fuzzy similarity for non-DOI entries; iii) constructs unified fields for keywords, abstracts and titles and merges keyword lists; iv) removes HTML artifacts, diacritics, punctuation inconsistencies and non-informative terms; v) parses cited references and extracts publication years for trend analyses.

The output is a validated and quality-checked metadata corpus that serves as the foundation for higher-layer analytics. Let us denote the bibliometric dataset that consist of ($N$) scientific publications indexed by $D_{raw} = \{r_1, r_2, \dots, r_N\}$. Each record ( $r_i$ ) is a structured tuple of WoS metadata fields $r_i = (T_i, A_i, K_i, F_i, Y_i, Abs_i)$ where: $T_i$ represents the title; $A_i = \{a_{i,1}, a_{i,2}, \dots, a_{i,m_i}\}$ is a finite set (or ordered list) of author names and ORCID tokens; $K_i = \{k_{i,1}, k_{i,2}, \dots, k_{i,p_i}\}$ represents the keywords as tokens in a controlled vocabulary $k_{i,l} \in \mathcal{K}$; $F_i = \{f_{i,1}, f_{i,2}, \dots, f_{i,q_i}\}$ represents the affiliations as tuples of institution and country; $Y_i$ is the publication year and $Abs_i$ is the abstract of the paper.

The full dataset can be represented as a set of tuples: $D_{raw} = \{(T_i, A_i, K_i, F_i, Y_i, Abs_i) | i = 1, \dots, N\}$. To extend analysis beyond metadata, the system incorporates a text enrichment pipeline that supports both abstract-level and full-text document analysis.

A textual subset $X_i$ is obtained by concatenating the title, authors, keywords and abstract. The following text enrichment methods are applied on $X_i$:

- Sentence-level and document-level embeddings ($e_i$) using transformer models (e.g., Sentence-BERT) are applied to support clustering, topic modeling, semantic search and research question expansion.

- Automated keywords expansion ($K_i^*$) using semantic similarity algorithms to identify additional relevant terms beyond those explicitly included in the metadata. It enhances topic detection and improves recall for user-defined thematic queries.
- Entity extraction (authors, topics, organizations, technical concepts) for downstream graph construction. Named-entity recognition (NER) modules extract key entities such as organizations, methods, technologies and research domains. Extracted entities ($\mathcal{E}_i$) later inform knowledge-graph visualizations and agent reasoning.

The final output of the layer is represented by $D_t = \{(r_i, e_i, K_i^*, \mathcal{E}_i, Y_i, Abs_i) | i = 1, \ldots, N\}$ that supports NL processing tasks to provide descriptive analytics, network construction, topic modeling (LDA, BERTopic) and agentic reasoning.

### 3.2 Analytics Layer

The analytics layer implements the full suite of scientometric and topic-modeling functionalities. It is subdivided into three modules as follows:

- Descriptive bibliometrics module performs classical analyses including publication growth trends, distribution of papers across journals and countries, co-authorship patterns, citation metrics and keyword frequency distributions.
- Network analytics module includes graph-based methods that are applied to analyze intellectual structures and collaboration patterns. Supported networks include co-authorship networks, institutional collaboration networks, keyword co-occurrence networks, co-citation and bibliographic coupling networks.
- Topic modeling module integrates both classical and modern topic-modeling approaches: Latent Dirichlet Allocation (LDA) and BERTopic for transformer-based, dynamic topic analysis. Dynamic topic evolution, hierarchical clustering of topics and generation of topic coherence metrics are included.

Further, let us define the Analytics Layer as a mapping $\mathcal{A}: D_t \mapsto A_t = (B_t, N_t, T_t)$, where $B_t$ denotes descriptive bibliometric indicators, $N_t$ denotes network-based analytical outputs and $T_t$ denotes topic-modeling results.

### 3.3 Agentic AI Layer

A core contribution of the framework is the incorporation of autonomous, goal-driven AI agents that coordinate complex tasks and enable NL interaction with the system. The layer consists of four specialized agents that interact as illustrated in the sequence diagrams in Supplementary material, section 1.1, Figures A1 and A2:

- *Custom Analytics Generator (CAG)* agent is based on NL prompts and interprets user requests such as "*plot the top 10 most cited papers from 2015 to 2020"* or *"detect emerging topics since 2021*". The AI agent performs intent detection, translation into internal analytic actions, code generation, code validation and safety checks, execution within an isolated computational cell. This enables non-technical users to perform complex analyses on the bibliometric dataset.
- *Paper Retriever (PR)* agent transforms research hypotheses and exploratory intents into enriched search queries and retrieves the top $n$ open-access full-text publications through the OpenAlex API. It downloads available PDFs of the open access papers, performs text extraction and forwards the processed corpus to the data ingestion, preprocessing and text enrichment layer. This capability transforms the platform from a metadata-based tool into a full-text bibliometric engine.
- *Research Assistant (RA)* performs higher-level interpretive tasks including research trend detection, topic gap identification and research hypotheses generation based on the outputs of the Analytics Layer and the evidence retrieved through the RAG pipeline. It acts as a "conceptual navigator", helping users explore the intellectual landscape of a field.
- *Report Generator (RG)* agent aggregates results, figures, tables and summaries into an exportable PDF or HTML report. It interprets the results using LLMs and enables users to rapidly generate research-ready bibliometric summaries.

The four-agent architecture is motivated by the fundamentally different cognitive and operational roles required in end-to-end bibliometric research as detailed in Table 1.

Table 1. The roles of the AI Agents

| Agent | Role | Requirements |
|---|---|---|
| **Custom Analytics Generator (CAG)** | Transform NL into executable scientometric code, validation, sandbox execution | Requires strict control, verification, safety constraints |
| **Paper Retriever (PR)** | External corpus expansion via OpenAlex, PDF ingestion | Requires external I/O, error handling, asynchronous retrieval |
| **Research Assistant (RA, RAG-based)** | Trend analysis, research gap identification and hypotheses generation | Requires grounded reasoning, hallucination control |
| **Report Generator (RG)** | Structured scientific narrative and document assembly | Requires long-context coherence and formatting |

The system employs a workflow orchestrator treated as a first-class architectural module that coordinates communication between agents, analytical modules and retrieval components. The orchestrator is not itself an autonomous reasoning agent but a deterministic control component that manages task scheduling, inter-agent communication, state propagation and execution flow. It operates on the user query $q_t$, the enriched bibliometric corpus $D_t$, knowledge base and the analytical outputs produced by the Analytics Layer.

Let the system state at step $t$ be defined as $S_t = (D_t, A_t, H_t, C_t, F_t, O_t)$, where $D_t$ denotes the current metadata and full-text corpus, $A_t$ denotes the outputs of the Analytics Layer, $H_t$ denotes the set of generated research hypotheses, $C_t$ denotes the set of retrieved text chunks or contextual evidence, $F_t$ denotes the FAISS vector index built over the available documents and $O_t$ denotes the collection of outputs produced during the session (figures, tables, summaries, reports and generated explanations). In this formulation, the orchestrator maintains a persistent analytical state and enables a closed-loop interaction in which bibliometric analytics generate hypotheses, hypotheses trigger literature retrieval and the newly retrieved evidence is reintegrated into the corpus for subsequent reasoning and analysis.

The workflow is formalized through the following sequence of state transitions.

1) The Analytics Layer first transforms the current corpus $D_t$ into structured bibliometric knowledge, including descriptive indicators, networks and topic structures. These outputs serve as the evidential substrate for the downstream agentic reasoning stage rather than being treated as terminal results.

$$A_t = \mathcal{A}_{AL}(D_t) \quad (1)$$

where $\mathcal{A}_{AL}(\cdot)$ denotes the Analytics Layer and $A_t = (B_t, N_t, T_t)$ comprises descriptive bibliometrics $B_t$, network-based structures $N_t$ and topic-modeling outputs $T_t$.

2) NL-to-code generation

$$O_t \leftarrow \mathcal{A}_{CAG}(q_t, A_t, O_t) \quad (2)$$

where $\mathcal{A}_{CAG}(\cdot)$ denotes the CAG agent.

3) Hypothesis generation from analytics

$$H_t \leftarrow \mathcal{A}_{RA}(q_t, A_t, O_t) \quad (3)$$

where $\mathcal{A}_{RA}(\cdot)$ denotes the RA. Given the user query $q_t$, the current analytical outputs $A_t$ and the session memory $O_t$, the agent identifies research gaps, synthesizes trends and generates a set of research hypotheses $H_t$.

4) Hypotheses transformed into retrieval queries

$$Q_t = T_{RA}(H_t) \quad (4)$$

where $T_{RA}(\cdot)$ denotes the query-generation function. It transforms the generated hypotheses into structured scholarly search queries $Q_t$, which are sent to OpenAlex to retrieve relevant open-access publications.

5) Corpus expansion after retrieval

$$D_t \leftarrow D_t \cup \mathcal{A}_{PR}(Q_t) \quad (5)$$

where $\mathcal{A}_{PR}(\cdot)$ denotes the document acquisition and preprocessing pipeline, including OpenAlex retrieval, PDF downloading when available, text extraction and metadata normalization. This step expands the corpus with newly retrieved papers, producing the updated document set $D_t$.

6) FAISS update and RAG context retrieval

$$F_t \leftarrow \mathcal{J}(D_{t+1}), C_t \leftarrow \mathcal{K}(q_t, H_t, F_{t+1}) \tag{6}$$

where $\mathcal{J}(\cdot)$denotes the embedding and FAISS indexing function, and $\mathcal{K}(\cdot)$ denotes the top-$k$ contextual retriever used in the RAG pipeline. The first operation updates the vector index over the expanded corpus, while the second retrieves the most relevant evidence chunks $C_{t+1}$ conditioned on the current user query and the generated hypotheses. The flowchart of the proposed framework is depicted in Figure 2.

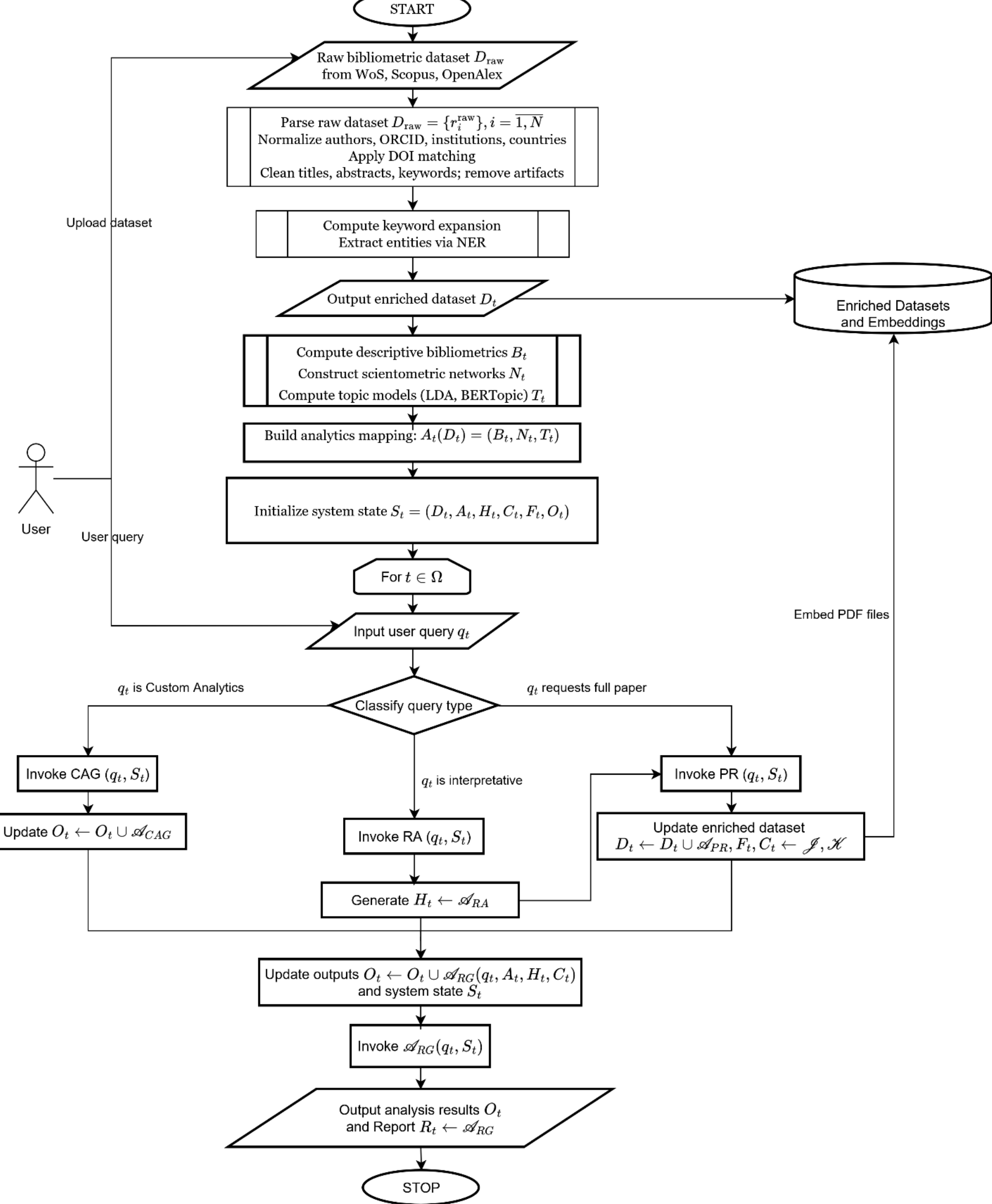

Figure 2. The flowchart of the proposed framework for AI-based bibliometric analysis

7) Grounded reasoning and report generation

$$O_t \leftarrow O_t \cup \mathcal{A}_{RG}(q_t, A_t, H_t, C_t) \quad (7)$$

where $\mathcal{A}_{RG}(\cdot)$ denotes the grounded response generation stage, which includes RAG-based interpretation by the RA and, when requested, structured synthesis by the RG. The resulting outputs may include textual explanations, identified gaps, refined hypotheses, comparative methodological analyses, figures, tables and exportable reports. Because $O_t$ is reinjected into the session state, the framework supports iterative, closed-loop exploration in which each reasoning cycle can trigger further retrieval, analysis and corpus expansion.

Accordingly, the proposed framework is not a linear pipeline but an iterative discovery system. The Analytics Layer produces structured scientometric evidence, the Agentic AI Layer converts this evidence into hypotheses and retrieval actions, and the retrieved literature is reintegrated into the corpus and FAISS index for subsequent RAG-based reasoning. The recursive design allows the system to refine its understanding of a scientific field and to support advanced, hypothesis-driven bibliometric investigation.

## 4. Simulations

To assess the proposed framework, two datasets are extracted from WoS: a dataset on bibliometric analyses and a dataset on energy communities. The first literature search on bibliometric research was conducted in WoS at the end of 2025. The search was executed using the Topic field with the query TS = ("bibliometric"). The results were filtered to include only articles and review articles, written in English and published between 2016 and 2025. Further restrictions were applied to include publications from major academic publishers (Elsevier, Springer Nature, MDPI, Frontiers, Emerald, Taylor & Francis, Wiley, Sage, IEEE) and indexed in Science Citation Index Expanded (SCI-EXPANDED) or Social Sciences Citation Index (SSCI). This systematic search returned a total of 20,794 documents, representing the current global research output on bibliometric methods, tools and applications across scientific fields. These records form the dataset used for subsequent bibliometric mapping, trend detection and AI-augmented analyses in this study. The searching formula of the first dataset is: TS = ("bibliometric") AND DT = ("Article" OR "Review Article") AND LA = ("English") AND PY = (2016÷2025) AND PU = ("Elsevier" OR "Springer Nature" OR "MDPI" OR "Frontiers Media" OR "Emerald Group Publishing" OR "Taylor & Francis" OR "Wiley" OR "Sage" OR "IEEE") AND WC = (SCI-EXPANDED OR SSCI).

The second literature search was conducted in the WoS to identify scientific publications related to the concept of energy communities. The search was performed using the Topic field (TS), which scans titles, abstracts and author keywords. The following query was applied: TS = ("energy community" OR "energy communities"), with no additional filters. This search yielded a total of 2,891 documents, representing the current body of peer-reviewed research addressing distributed energy models, renewable energy communities and local energy governance. The searching formula for the second dataset is: TS = ("energy community" OR "energy communities"). These results constitute the dataset foundation [2] for the bibliometric and AI-augmented analyses conducted in our research.

### 4.1 Process datasets with data and analytics layers

#### 4.1.1 Bibliometric dataset

The comprehensive bibliometric dataset comprises 20,768 publications indexed between 2016 and 2025, totaling 508,561 citations, 50,884 unique authors and 36,615 unique keywords. Before conducting descriptive and scientometric analyses, the dataset undergoes the multi-step processing pipeline of the data layer. First, raw records are parsed and standardized, including normalization of author names, ORCID identifiers, institutional affiliations and country metadata. Metadata completeness is improved using DOI resolution and fuzzy matching techniques. Titles, abstracts and keywords are cleaned to remove noise and formatting artifacts. From each record, structured fields are extracted (titles, authors, keywords, funding information, publication year and abstracts), forming a unified textual representation for downstream analysis.

[2] https://github.com/AdelaBara/AI-Research-Lens

To enrich the textual layer, embeddings are generated for each document, extended keyword sets are created through keyword expansion techniques and named entities are extracted using NER models. These operations yield the enriched dataset $\widetilde{D}$ integrating both structured metadata and semantic representations. This dataset is subsequently processed by the analytics layer to compute the descriptive bibliometric indicators, including annual publication counts, citation distributions, top-cited articles, most productive authors and institutions, keyword growth rates and heatmaps of keyword usage over time. Publication and citation trends show sustained growth throughout the period, peaking in 2024, with a strong increase in scientific output from 377 in 2016 to 4675 in 2025. The corpus is multidisciplinary, spanning major areas such as Environmental Sciences, Engineering, Computer Science, Business and Public Health (Figure 3a). Citation activity is particularly concentrated in recent years (Figure 3b), with 2021 exhibiting the highest annual citation count (93048 citations), reflecting the field's accelerating scholarly impact. The presence of highly influential works is notable: the most cited article alone has accumulated more than 9,600 citations, while several others surpass the 3,000-5,000 citation range, highlighting the dominance of methodological and review-oriented contributions.

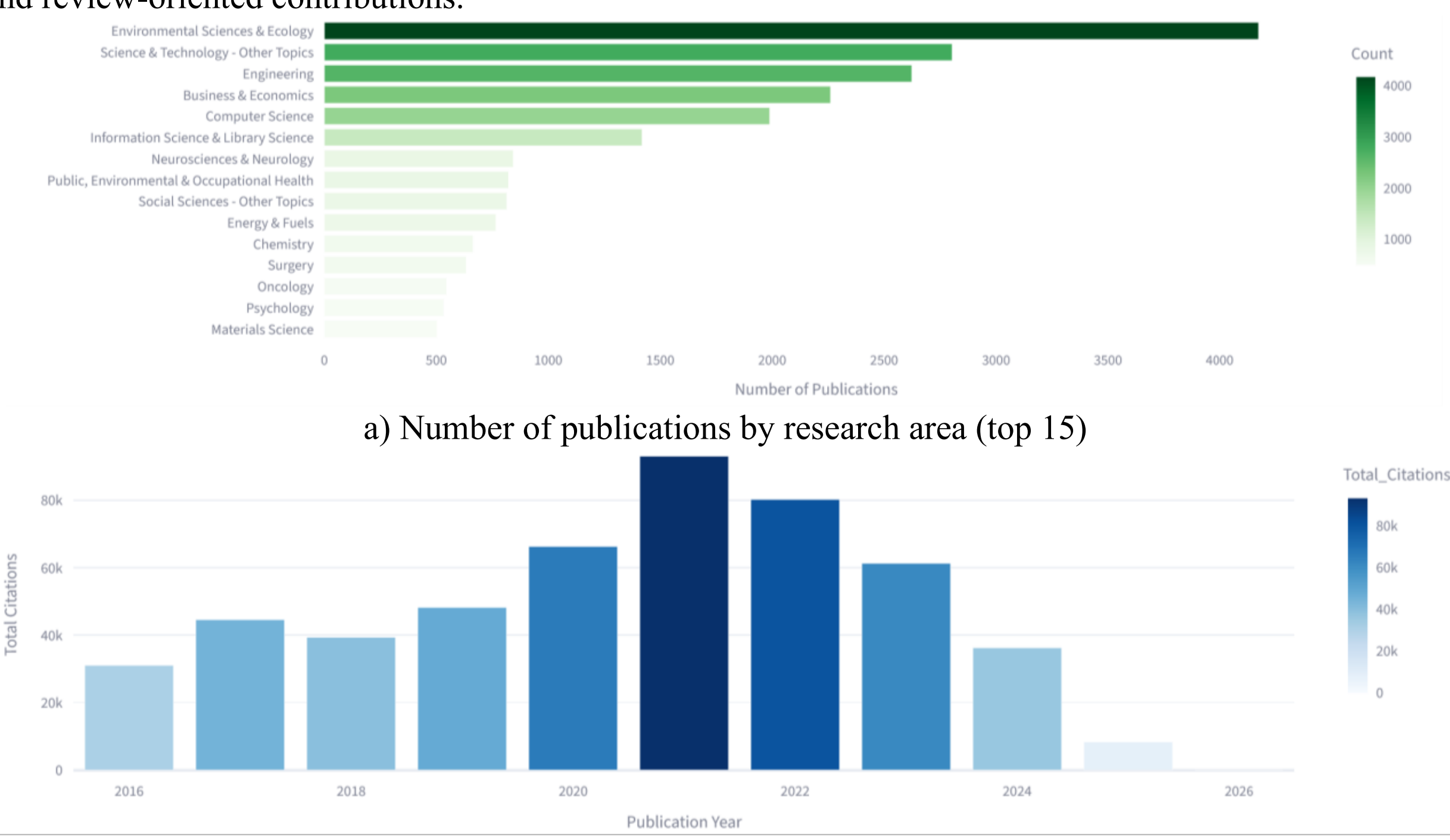

a) Number of publications by research area (top 15)

b) Number of citations per year

Figure 3. Number of publications and citations between 2016 and 2025

Keyword statistics further underscore the field's dynamism, with a vocabulary of 36,615 unique terms and strong growth observed in high-momentum topics such as bibliometrics, science mapping, VOSviewer, CiteSpace, systematic review and emerging themes such as sustainability and artificial intelligence. The keyword co-occurrence network in Figure 4 highlights the conceptual structure of the research domain and reveals how frequently major themes are linked within the corpus. Bibliometric analysis emerges as the central pivot of the network, exhibiting the highest connectivity and the strongest link weights. Tools and methodological platforms such as VOSviewer and CiteSpace also occupy structurally important positions, with strong ties (e.g., the link between CiteSpace and bibliometric reaches 200, while its connection to VOSviewer stands at 113). Emerging thematic areas show high integration with core bibliometric concepts: artificial intelligence connects with weights ranging from 12 to 24, indicating a steady but developing intersection between AI applications and bibliometric workflows. Similarly, sustainability displays moderate but growing link strengths (e.g., 22, 15), reflecting its increasing presence as a topic of inquiry within the bibliometric literature. More established scholarly practices such as literature review and

research trends are also well integrated, with multiple connections in the 10-70 range, suggesting their continued methodological relevance.

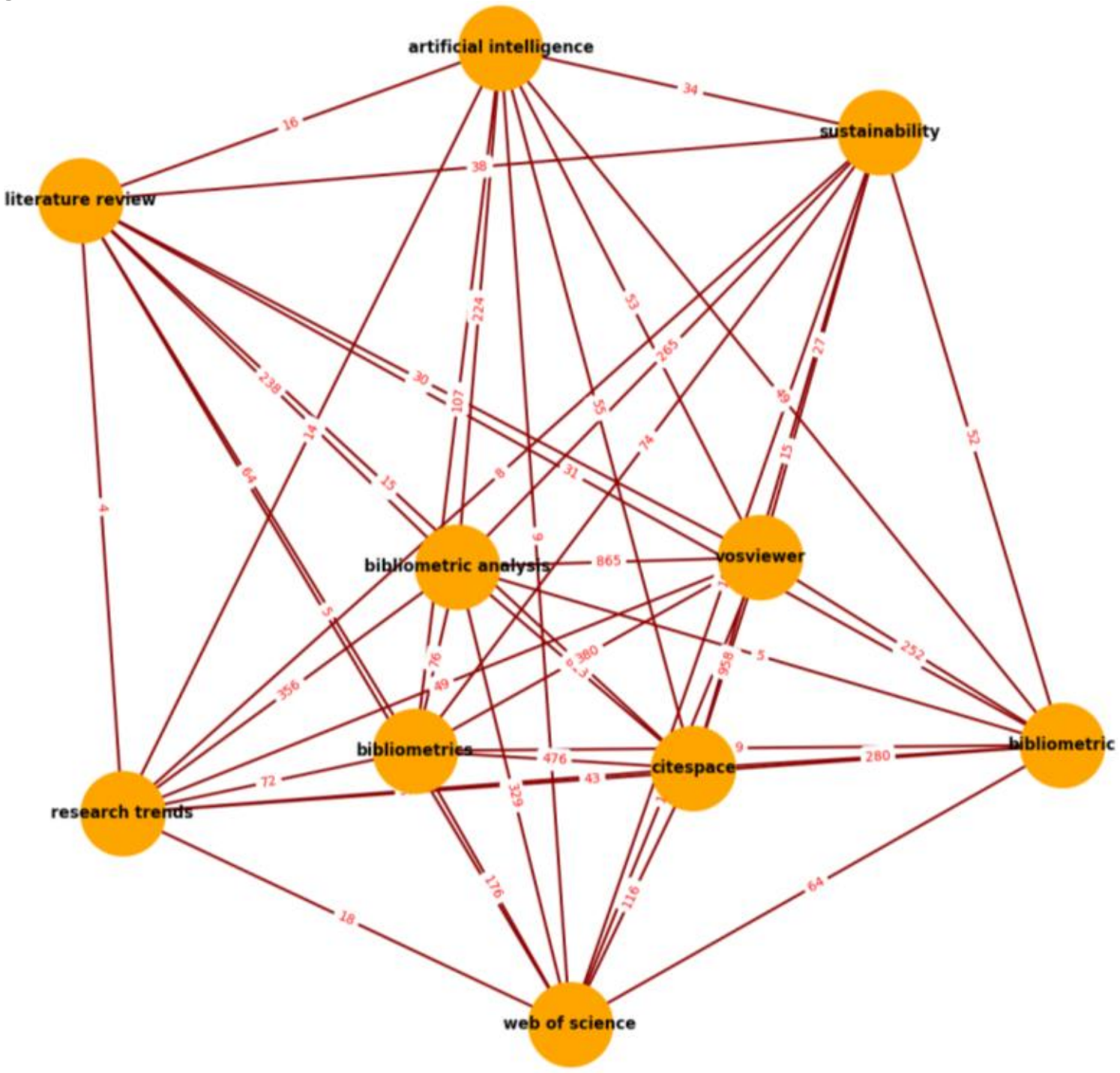

Figure 4. Keywords analysis of the bibliometric dataset

### 4.1.2 Energy community-related dataset

For the energy community dataset, the corpus comprises 2,891 publications, with a total of 49,530 citations and contributions from 7,447 unique authors. While early output remained modest for several decades, publication activity accelerates sharply after 2018, culminating in a pronounced peak around 2023-2024, consistent with the rising policy and scientific interest in community energy systems. Keyword statistics confirm the thematic consolidation of the field: among 6,535 unique keywords, terms such as energy community, renewable energy, energy transition and energy sharing dominate the corpus (Figure 5a), reflecting the centrality of decentralization, prosumerism and renewable integration. The most cited works exceed 1,000-2,300 citations, indicating the presence of highly influential conceptual and policy-oriented papers. Disciplinary distribution in Figure 5b shows that Energy & Fuels, Science & Technology, Environmental Sciences & Ecology, Engineering and Business & Economics together account for the largest proportion of publications, underscoring the field's interdisciplinary nature and its strong alignment with sustainability and energy transition research.

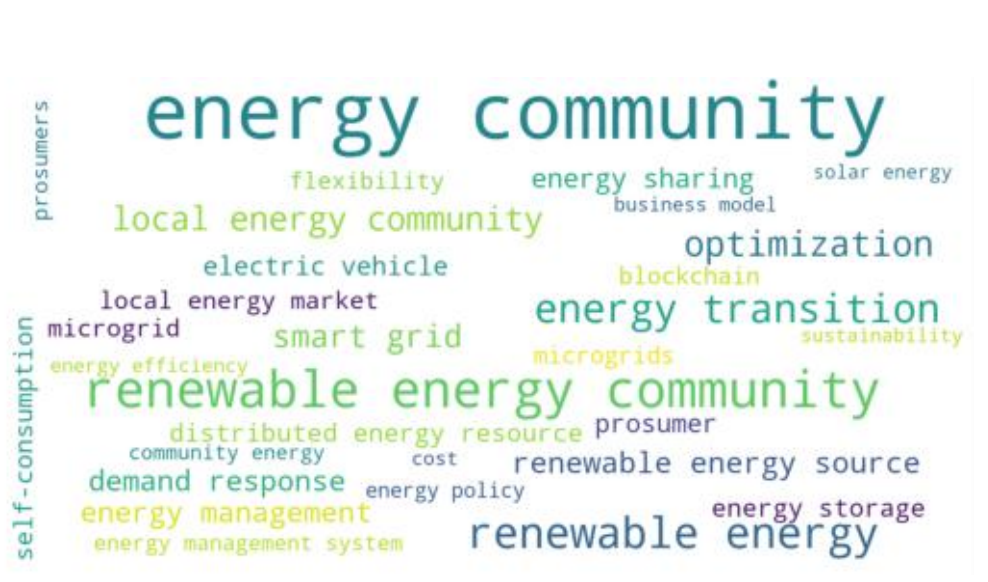

a) Word cloud

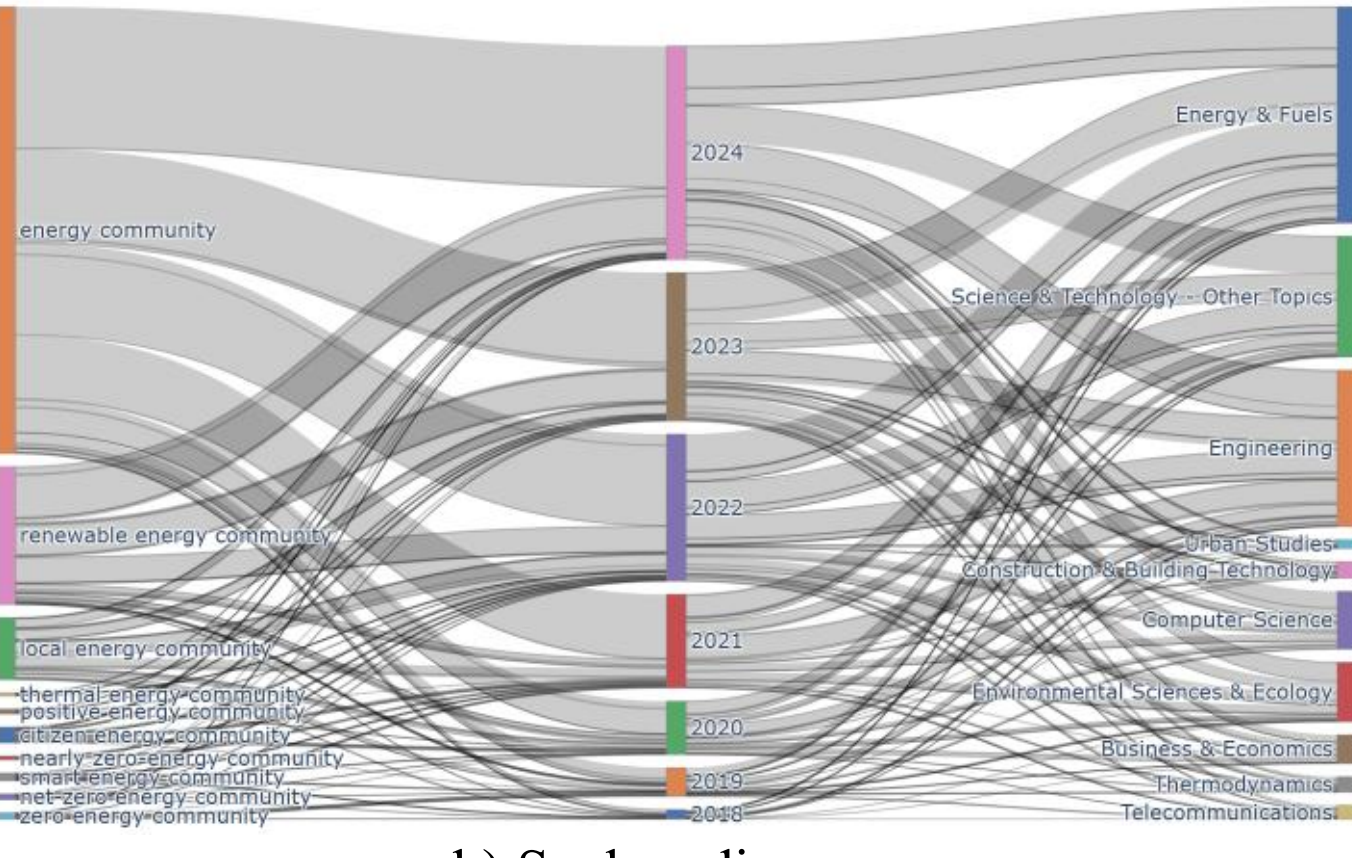

b) Sankey diagram

Figure 5. Keywords analytics for the energy community dataset

For the topic modeling simulations, BERTopic is applied with tunable UMAP and HDBSCAN parameters to control the granularity of the resulting topic structure. UMAP is used to generate a lower-dimensional semantic representation of the enriched textual dataset, with the number of neighbors and components determining the level of local versus global structure preserved in the embedding. Lower values of *n_neighbors* and *n_components* promote more fine-grained semantic separation and therefore facilitate the emergence of a larger number of clusters. The clustering stage is performed using HDBSCAN, where the *min_samples* parameter controls the minimum cluster density and the *cluster_selection_epsilon* influences the tendency of the algorithm to split or merge clusters. BERTopic is configured with the following parameters: *min_topic_size = 20, n_gram_range = (1, 2), top_n_words = 20, embedding_model = all-MiniLM-L6-v2*. Dimensionality reduction is performed using UMAP with parameters: *n_neighbors = 15, n_components = 5, metric = "cosine", min_dist = 0.0, random_state = 42*. Clustering is performed using HDBSCAN with parameters *min_samples = 10, metric = "euclidean", cluster_selection_method = "eom", cluster_selection_epsilon = 0.5*. The vectorizer uses *min_df = 2, max_df = 0.95* and English stopwords. These values follow standard BERTopic recommendations and are selected to achieve a balance between topic coherence, noise suppression and prevention of overly fragmented topic structures.

The BERTopic results reveal clear clustering patterns emerging across both the UMAP (Figure 6) and t-SNE embeddings. The largest and most densely populated cluster corresponds to Topic 0, which contains 1,936 documents and is associated with the broad theme of Renewable & Renewable Energy & Storage, indicating the dominant position of renewable energy research within the corpus. Several smaller but thematically distinct clusters also appear: Topic 1 (160 documents) centers on Blockchain & Smart/IoT applications, Topic 2 (157 documents) focuses on Learning, Forecasting and Load Modeling, while Topic 3 (104 documents) groups literature on Nuclear, Euratom and Atomic Energy. Additional clusters capture specialized subfields such as Wind & Turbine Dynamics (65 documents), Wave & Marine Energy (36 documents) and Solar Materials & Cells (27 documents). Topic -1 (308 documents) are considered as Miscellaneous/Outliers, reflecting heterogeneous or weakly connected texts that do not align strongly with any major theme. The intertopic distance map shows that these clusters are well separated, with minimal overlap, confirming that BERTopic successfully differentiated distinct domains of research activity.

**4.2 Agentic AI framework and assessment**

The agentic AI framework is implemented through a modular stack in which a task-aware orchestrator coordinates multiple domain-specific components. The system integrates a prompt-driven code generator that iteratively refines outputs through a feedback loop, a FAISS-based vector retrieval layer and a *Gemini 3 Flash* model optimized for RAG. Inference is configured with *temperature* = 1 and *max_tokens* = 3000. No explicit *top-p* (nucleus sampling) parameter is set and the provider's default decoding policy is applied. The temperature value is set to 1 to balance response diversity and reasoning flexibility when handling complex analytical tasks. Compared with earlier model generations, Gemini 3 is designed to operate effectively at higher temperature settings, allowing the model to explore alternative reasoning paths and produce more expressive analytical explanations while remaining grounded in the retrieved context. The retrieval layer is implemented using FAISS over dense embeddings generated by *text-embedding-3-small* with an embedding dimensionality of 1536. Retrieved contexts are then fused into the LLM prompt, enabling grounded, domain-aware and highly responsive analytic workflows.

The *Gemini 3 Flash* model is employed as an inference-only backbone for AI Agents. It is responsible for NL-to-code generation, structured Python code synthesis and iterative correction under sandboxed execution constraints. The same backbone performs retrieval-augmented reasoning over full-text evidence retrieved from the FAISS index, enabling grounded interpretation, trend analysis and research hypotheses generation.

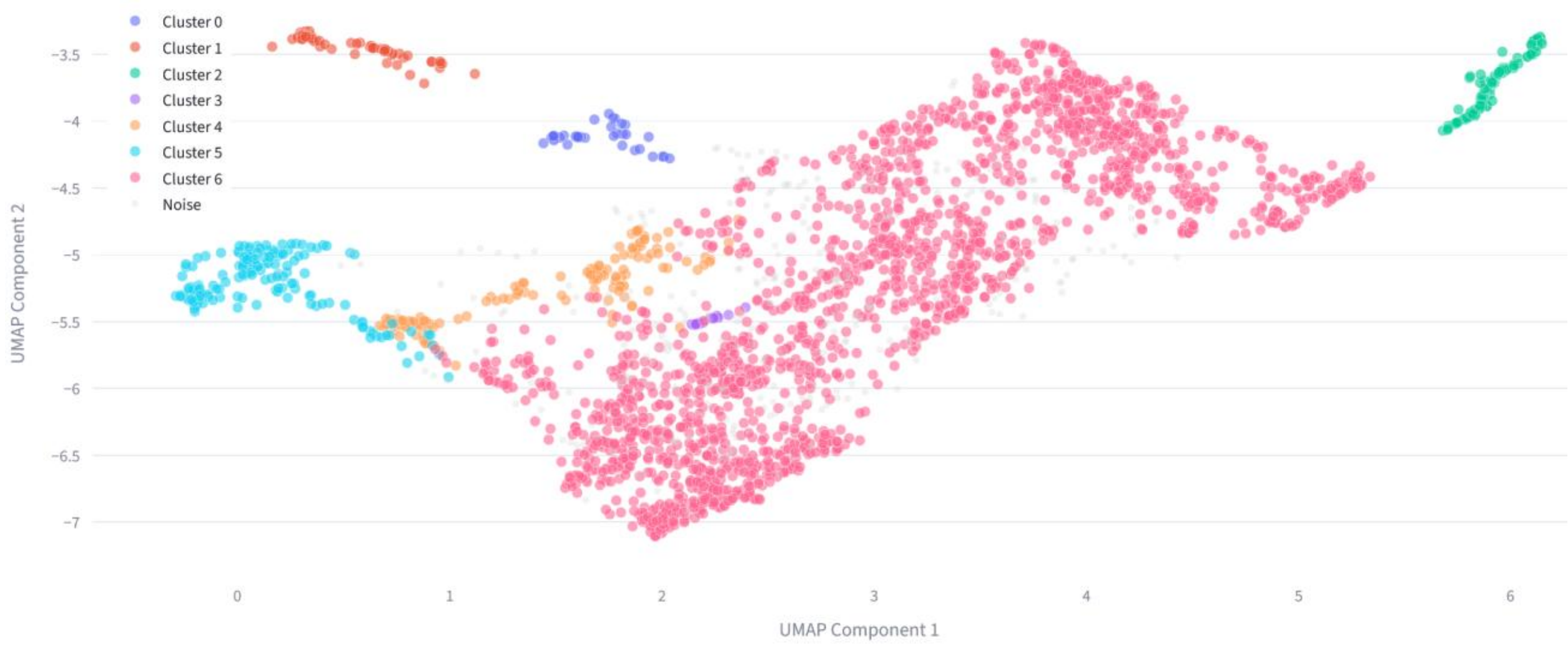

Figure 6. Cluster distribution in UMAP space for energy community dataset

To evaluate the CAG AI agent which automatically generates and executes analytic code in response to NL prompts, a human evaluation involving researchers with prior experience in energy community systems, local electricity markets and energy policy analysis was performed. It consists of a multi-stage assessment test on 200 prompts focusing on correctness, quality and alignment with user intent. These prompts were collected from two sources: (i) real exploratory queries from graduate students and domain researchers conducting bibliometric studies, (ii) scenario-driven prompts created by the authors to emulate standard scientometric workflows. The prompts were grouped into several task categories, including topic discovery, thematic analysis, methodological comparison and domain-specific research questions. To capture varying levels of analytical difficulty, the prompts were constructed with different degrees of complexity, ranging from simple information retrieval requests to more elaborate queries requiring synthesis across multiple documents or research themes. The prompts range from 10 to ~60 words, depending on task complexity and include both analytical queries handled by the RA and CAG. For CAG, examples of prompts include: *"What are the main research themes in the literature on energy communities and local electricity markets?", "How has research on peer-to-peer energy trading in energy communities evolved over time?", "Show publication trends by year using a line chart.", "Produce a histogram showing the distribution of publications across research categories.", "Visualize the most frequent keywords using a frequency chart"*. This prompt dataset is representative of the system's intended users: domain researchers with limited programming expertise who express complex analytical goals through NL. Samples of the prompts and outputs of AI-Research-Lens app are illustrated in Figure 7.

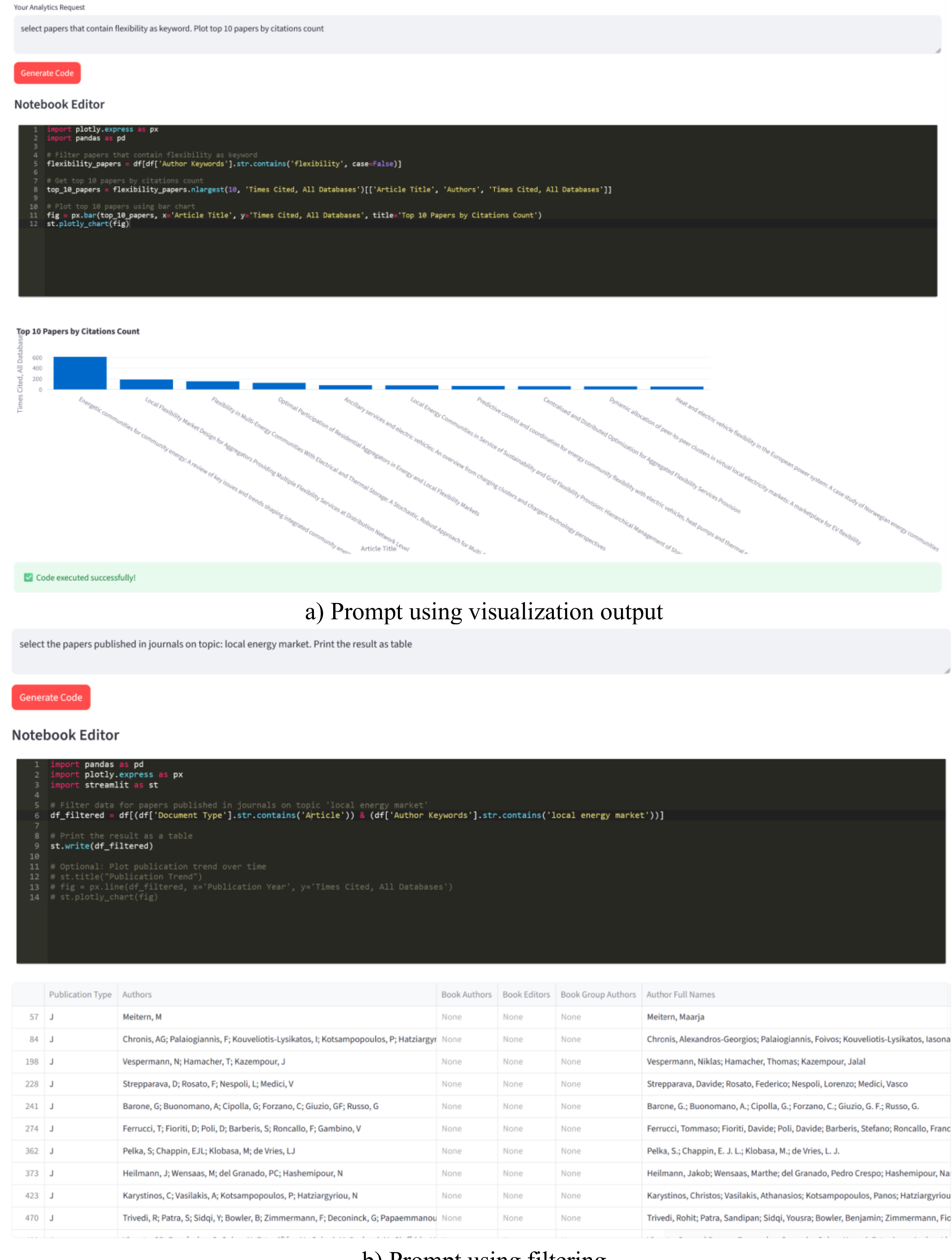

|  | Publication Type | Authors | Book Authors | Book Editors | Book Group Authors | Author Full Names |
|---|---|---|---|---|---|---|
| 57 | J | Meitern, M | None | None | None | Meitern, Maarja |
| 84 | J | Chronis, AG; Palaiogiannis, F; Kouveliotis-Lysikatos, I; Kotsampopoulos, P; Hatziargyr | None | None | None | Chronis, Alexandros-Georgios; Palaiogiannis, Foivos; Kouveliotis-Lysikatos, Iasona |
| 198 | J | Vespermann, N; Hamacher, T; Kazempour, J | None | None | None | Vespermann, Niklas; Hamacher, Thomas; Kazempour, Jalal |
| 228 | J | Strepparava, D; Rosato, F; Nespoli, L; Medici, V | None | None | None | Strepparava, Davide; Rosato, Federico; Nespoli, Lorenzo; Medici, Vasco |
| 241 | J | Barone, G; Buonomano, A; Cipolla, G; Forzano, C; Giuzio, GF; Russo, G | None | None | None | Barone, G.; Buonomano, A.; Cipolla, G.; Forzano, C.; Giuzio, G. F.; Russo, G. |
| 274 | J | Ferrucci, T; Fioriti, D; Poli, D; Barberis, S; Roncallo, F; Gambino, V | None | None | None | Ferrucci, Tommaso; Fioriti, Davide; Poli, Davide; Barberis, Stefano; Roncallo, Franc |
| 362 | J | Pelka, S; Chappin, EJL; Klobasa, M; de Vries, LJ | None | None | None | Pelka, S.; Chappin, E. J. L.; Klobasa, M.; de Vries, L. J. |
| 373 | J | Heilmann, J; Wensaas, M; del Granado, PC; Hashemipour, N | None | None | None | Heilmann, Jakob; Wensaas, Marthe; del Granado, Pedro Crespo; Hashemipour, Na |
| 423 | J | Karystinos, C; Vasilakis, A; Kotsampopoulos, P; Hatziargyriou, N | None | None | None | Karystinos, Christos; Vasilakis, Athanasios; Kotsampopoulos, Panos; Hatziargyriou |
| 470 | J | Trivedi, R; Patra, S; Sidqi, Y; Bowler, B; Zimmermann, F; Deconinck, G; Papaemmanou | None | None | None | Trivedi, Rohit; Patra, Sandipan; Sidqi, Yousra; Bowler, Benjamin; Zimmermann, Fic |

a) Prompt using visualization output

b) Prompt using filtering

Figure 7. Sample prompts and output of the CAG AI agent

The syntactic correctness was verified by checking that generated code executed without errors and interacted properly with the underlying dataset. The semantic correctness was assessed by validating that the outputs (e.g., tables, charts, statistics) matched ground-truth computations performed manually or through independent scripts. For example, when prompted to "*create an interactive horizontal bar chart of* *the top 20 keywords*", we verified that exactly 20 items were displayed, that keyword frequencies corresponded to the dataset and that the visual structure adhered to the requested specifications.

Also, consistency was examined by re-issuing identical prompts and ensuring reproducible outputs. We evaluated user-intent fidelity by inspecting whether the generated code conformed precisely to the prompt, neither adding unnecessary operations nor omitting required elements. All reported human evaluation scores were obtained from seven independent expert raters with doctoral-level training in data science, scientometrics and AI. Each rater independently assessed system outputs using a standardized rubric across four dimensions: syntactic correctness, semantic validity, prompt adherence and output quality, employing a 1–10 Likert scale. To assess the consistency of human judgments, we computed Krippendorff's alpha ($\alpha = 0.82$), indicating strong inter-rater agreement and high reliability of the evaluation results.

To complement human evaluation, we used GPT-5.2 as an LLM-as-a-judge. The model reviewed the generated code and outputs, scoring them on syntactic correctness, semantic validity and visualization quality. The LLM-as-a-judge was employed as a complementary evaluation mechanism to examine convergence with human assessment. Its scores are not used to replace human judgment but to provide an additional validation signal. The close alignment observed between automated and human ratings further supports the robustness of the reported performance metrics. The results are centralized in Table 2.

Table 2. Evaluation metrics of the CAG

| Metric | Description | Scores |
|---|---|---|
| **Syntactic accuracy (%)** | Code blocks with no syntax errors | 95% |
| **Semantic accuracy (%)** | Outputs producing correct values/logic | 90% |
| **Prompt alignment score (1–10)** | Mean alignment with user instructions | 9.2 |
| **Output quality score (1-10)** | Average quality of charts/tables | 9.5 |
| **Reproducibility (%)** | Prompts yielding consistent results across multiple runs | 94% |
| **LLM-judge mean score (1–10)** | GPT-5.1 evaluation of correctness and clarity | 9.3 |
| **Human mean score (1–10)** | Human rating of correctness and clarity | 9.2 |

The model achieved a syntactic accuracy of 95%, indicating that nearly all generated code executed without errors and was structurally valid. Semantic accuracy reached 90%, meaning that in most cases the output (such as tables, plots or computed statistics) reflected correct logic and values when compared to ground-truth computations. The model is aligned with user intent, with an average prompt alignment score of 9.2 and produces high-quality visual and tabular outputs, achieving an output quality score of 9.5.

Robustness was also high, with 94% reproducibility, confirming that repeated prompts generated consistent responses. The LLM-as-a-judge evaluation, performed by GPT-5.1, produced a mean score of 9.3, closely matching the human evaluation average of 9.2, indicating strong agreement between human experts and automated assessment. Further, the results are divided by the following type of prompts: visualization requests, data filtering and selection and statistics (as in Table 3). Most of the prompts include at least 2 types (for e.g., filtering + visualization or statistics + visualizations)

Table 3. CAG agent evaluation results divided by prompt types

| Prompt type | Prompts | Syntactic accuracy (%) | Semantic accuracy (%) | Output quality score (1-10) |
|---|---|---|---|---|
| **Visualization** | 130 | 98% | 92% | 9.8 |
| **Data filtering** | 180 | 98% | 94% | 9.5 |
| **Statistics** | 150 | 93% | 87% | 8.8 |

Visualization prompts that involved plotting the results achieved the highest scores overall, with 98% syntactic accuracy, 92% semantic accuracy and an output quality score of 9.8, reflecting the model's strong capability in generating correct and aesthetically coherent graphical outputs. Data-filtering prompts also showed high performance, with 98% syntactic and 94% semantic accuracy, demonstrating the system's reliability in handling dataset queries, selections and transformations on the dataset. Statistical prompts, while still performing well, exhibited slightly lower scores, with 93% syntactic accuracy, 87% semantic accuracy and an output quality score of 8.8. It suggests that analytic computations (such as aggregations, descriptive statistics or computed indicators) introduce increased complexity and therefore higher error potential. Nevertheless, all three categories maintain high performance levels, confirming that the CAG agent generalizes effectively across diverse prompt types, with only minor degradation on more numerically intensive tasks.

The responses of the RA agent are evaluated across a set of established retrieval and generation metrics widely used in RAG research. The selected metrics capture both the quality of the retrieval stage (e.g., Recall@5, Mean Reciprocal Rank-MRR) and the accuracy, grounding and usefulness of the responses (e.g., faithfulness, correctness, relevance, hallucination rate). In addition, end-to-end evaluation criteria such as human ratings, context utilization and latency provide insight into the responsiveness of the pipeline. For the evaluation of the RA, we similarly curated 200 prompts reflecting RA capabilities to retrieve evidence from multiple documents, perform cross-paper reasoning and generate grounded hypotheses. Table 4 centralizes the scores obtained by interactions between user and RA agent.

Table 4. Evaluation metrics of the RA agent

| Metric Category | Metric | Description | Score |
|---|---|---|---|
| **Retrieval quality** | **Recall@5** | Fraction of queries for which at least one relevant document appears in the top-5 retrieved results. Measures how well FAISS retrieves necessary evidence. | 0.93 |
| | **MRR** | Evaluates how early the correct document appears in the ranking; higher values mean relevant content is retrieved sooner. | 0.87 |
| **Generation quality** | **Faithfulness** | Assesses whether the generated response stays grounded in the retrieved context without hallucinations. | 9.1/10 |
| | **Correctness** | Measures factual accuracy of the output relative to ground-truth information in the documents. | 8.9/10 |
| | **Hallucination rate** | Percentage of responses that contain unsupported or fabricated claims (lower is better). | 5% |
| | **Relevance** | Evaluates alignment between the generated answer and the user question; higher scores indicate more directly useful responses. | 9.4/10 |
| **RAG performance** | **Human rating** | Human overall rating of clarity, usefulness and trustworthiness of the answers. | 9.2/10 |
| | **Context utilization ratio** | Share of retrieved context chunks that were referenced by the model. Indicates efficiency of retrieval. | 81% |
| | **Latency** | Average total system response time (retrieval + generation). Useful for assessing real-time usability. | 5.4 s |

Retrieval performance achieved a Recall@5 of 0.93 and an MRR of 0.87, demonstrating FAISS reliably in retrieving relevant documents early in the ranking. Generation metrics further confirm that the model produces accurate responses: faithfulness reaches 9.1/10, correctness is 8.9/10 and hallucinations remain low at 5%, while relevance achieves 9.4/10, showing close alignment with user questions. End-to-end evaluation reflects high performance, having a human rating of 9.2/10 and efficient use of retrieved context, as evidenced by an 81% context utilization ratio. Although the average latency of 5.4 seconds is higher than real-time systems, it remains acceptable for research-oriented interactions requiring retrieval and grounding. Figure A6 in Supplementary material illustrates 3 samples of conversations between user and the RA agent regarding the content of the ingested papers.

The RG AI agent serves as the final synthesis layer of the system, aggregating all outputs produced across the analytics pipeline, including descriptive statistics, topic models, network structures, contextual retrieval and code-generated visualizations, into a coherent, automatically assembled document. Leveraging the results generated by the CAG and RA agents and the underlying bibliometric analyses, the RG agent composes structured narratives that combine textual interpretation with dynamically produced figures and tables. At runtime, it integrates the most relevant analytical outputs, inserts context-aware commentary and formats the complete report into a user-exportable PDF or HTML document, enabling users to obtain full analytical reports without manual assembly or post-processing.

### 4.3 AI-enabled insight discovery beyond traditional bibliometrics

While the preceding analyses present standard descriptive and scientometric results, the principal contribution of the proposed framework lies in its ability to generates research hypotheses through an agent-driven exploration that cannot be obtained through conventional bibliometric tools.

In the first case study using the bibliometric dataset, the system linked topic evolution with citation trajectories and full-text interpretation, revealing a growing methodological divide between computational bibliometrics (AI-driven topic modeling, embeddings, deep learning) and policy-driven scientometrics, with limited methodological cross-fertilization between these items. The framework further identified underexplored research gaps at this intersection, suggesting opportunities for hybrid evaluation methodologies that combine algorithmic modeling with policy-oriented impact assessment. These insights emerge from the system's ability to dynamically generate hybrid metrics, integrate full-text semantic reasoning and iteratively refine analysis pipelines via NL interaction, capabilities that are not supported by existing bibliometric software, which relies on static, metadata-only workflows.

In the second case study using the energy communities' dataset, using NL prompts, the system dynamically composed a multi-stage analytical pipeline combining temporal citation bursts, BERTopic topic evolution, full-text semantic extraction from retrieved articles and RAG-based synthesis. This analysis revealed that blockchain-related research in energy communities initially emerged within high-impact engineering journals but subsequently migrated toward policy-oriented and interdisciplinary venues after 2020, coinciding with the introduction of national and EU regulatory frameworks for local energy markets. Moreover, by extracting evidence from the discussion sections of retrieved full-text papers, the RA identified regulatory uncertainty and interoperability constraints as the primary bottlenecks limiting large-scale deployment of energy communities, a conclusion that does not surface from keyword co-occurrence networks or citation maps alone and would otherwise require extensive manual literature review. Using the proposed RG with domain-aware interpretation prompts, the system automatically produced several non-obvious hypotheses on the energy communities' corpus. Among these, it detected a structural anomaly: the paper "MadGraph 5: Going Beyond", a high-energy physics software article, appeared as the most cited publication in the energy communities' dataset (2,655 citations), despite having low connection to the domain's dominant keywords (e.g., energy community, renewable energy community, energy transition). The system further uncovered cross-layer interdependencies between technical optimization research (e.g., smart grids, energy management, demand response) and socio-economic policy frameworks (e.g., peer-to-peer markets, EU Clean Energy Package), revealing energy communities as an inherently sociotechnical system. It also identified latent research gaps: despite the central role of the term "community", social, behavioral and equity dimensions are strongly underrepresented in high-frequency keywords, suggesting a systematic imbalance in current research. These insights are further reported by the RG through dynamic metric synthesis, cross-layer reasoning over bibliometric structure and semantic interpretation of retrieved texts, capabilities that are difficult to obtain using traditional metadata-only bibliometric tools without extensive manual analysis. The output of the RG is provided in Supplementary material (Figure A3 and Figure A4). To illustrate the system's support for interactive, iterative refinement, we present a representative multi-round analytical session on the energy communities' dataset in Supplementary material in section 1.4. Also, to demonstrate the system's capability to dynamically create new evaluation metrics, we present an end-to-end example from the energy communities case study in Supplementary material, in section 1.5.

### 4.4. Ablation study

To justify the necessity of the proposed four-agent architecture, we conducted an ablation study comparing the full system against a streamlined baseline configuration implemented as a single LLM agent equipped with tool calling for code generation, retrieval and reporting. The ablation study is conducted on a curated evaluation set consisting of 200 prompts per agent component, designed to cover representative analytical tasks including literature exploration, thematic analysis and code-based bibliometric computation, covering tasks handled by both the RA and CAG. The prompt dataset was balanced across analytical task categories to ensure comparable coverage of retrieval, analytical reasoning and code-execution scenarios.

Uncertainty in the evaluation results was quantified using several statistical procedures. For binary outcomes, including syntactic execution success, reproducibility and hallucination incidence, 95% confidence intervals were computed using Wilson binomial intervals based on the prompt-level outcomes. For continuous evaluation scores assigned by human reviewers and LLM-based judges (on a 1–10 scale), 95% confidence intervals for the mean were estimated using nonparametric bootstrap resampling with 1,000 resamples.

Architectural comparisons between the single-agent baseline and the four-agent system were conducted using the same prompt set, allowing paired statistical testing. Differences between configurations were evaluated using paired Wilcoxon signed-rank tests with a significance threshold of $\alpha = 0.05$.

The results in Table 5 indicate that the multi-agent architecture improves stability and reliability across all evaluation dimensions. The single-agent configuration exhibited lower robustness, with more frequent error propagation across tasks such as code execution, retrieval and interpretation, as well as higher hallucination incidence during semantic reasoning. In contrast, the proposed four-agent framework achieved 95% code execution success, 90% semantic accuracy, 94% reproducibility and a 5% hallucination rate, while the single-agent baseline achieved 81% execution success, 76% semantic accuracy, 78% reproducibility and a 14% hallucination rate.

Table 5. Results of the ablation study comparing the single-agent baseline with the proposed four-agent architecture across key evaluation metrics.

| Evaluation Metric | Single-Agent Baseline | Four-Agent Architecture | 95% Confidence Interval |
|---|---|---|---|
| **Code execution success** | 81% | 95% | Wilson CI (binomial) |
| **Semantic accuracy** | 76% | 90% | Bootstrap CI (1,000 resamples) |
| **Reproducibility** | 78% | 94% | Wilson CI (binomial) |
| **Hallucination rate** | 14% | 5% | Wilson CI (binomial) |

These results show that separating responsibilities across specialized agents improves controllability and robustness in complex AI-driven analytical workflows. Therefore, the architecture reduces task interference and improves the reliability of end-to-end scientometric analysis.

## 5. Conclusions

The paper proposes an integrated agentic AI framework designed to streamline bibliometric and scientometric inquiry. It combines RAG approach, FAISS-based semantic search, dynamic NL-to-code analytics and automated report synthesis to provide an adaptive and extensible research environment capable of supporting rich, data-driven exploration of scientific literature.

The architectural complexity of proposed framework arises from its integration of multiple interdependent agentic components, each performing specialized functions being coordinated by the orchestrator workflow. The framework operates across several abstraction layers (semantic retrieval, NL reasoning, code generation, sandboxed execution, topic modeling, network construction and automated reporting) which must interact dynamically to produce coherent outputs. This multi-layer orchestration requires integrated state management, context propagation and adaptive decision-making, particularly as user queries vary in scope and intent. The system's reliance on heterogeneous models (dense embedding models, FAISS indexing, LLMs for RAG and execution engines for real-time code evaluation) further increases architectural heterogeneity and necessitates synchronization to avoid failure cascades. Despite this inherent complexity, the modular design enables each component to remain independently optimizable while contributing to a coordinated agentic AI system that can successfully manage intricate research workflows without imposing cognitive or technical overhead on the user.

The empirical evaluation demonstrates that the framework delivers high reliability across all components: the CAG agent achieved 95% syntactic accuracy, 90% semantic accuracy and strong alignment with user instructions, while the LLM-as-a-judge and human reviewers converged on highly consistent quality scores. Likewise, the RA agent that incorporates the RAG module exhibited a Recall@5 of 0.93 and MRR of 0.87 and generated grounded responses with minimal hallucination (5%) and high relevance (9.4/10).

Although the framework supports automated analytical workflows, the interpretation of scientometric outputs still requires domain expertise. The system should therefore be considered an AI-assisted research support tool that facilitates exploration and analysis of large scientific corpora rather than a fully autonomous scientific analysis system. Future research will include controlled benchmarking against traditional bibliometric workflows and existing AI-assisted research tools.

**Statements and Declarations**
**Ethical Approval.** Not applicable
**Consent to Participate.** Not applicable
**Consent to Publish.** Not applicable
**Authors Contributions.** Contribution to the study conception and design: SVO and AB. Material preparation, data collection and analysis were performed by SVO. The first draft of the manuscript was written by SVO and AB, the second draft by SVO, and they also commented on all versions of the manuscript. The authors read and approved the final manuscript.
**Disclosure statement/Competing Interests.** The authors report there are no competing interests to declare.
**Data availability statement.** The data is available on https://github.com/AdelaBara/AI-Research-Lens
**Acknowledgement and Funding**: This work was supported by a grant of the Ministry of Research, Innovation and Digitization, CNCS/CCCDI - UEFISCDI, project number COFUND-DUT-OPEN4CEC-1, within PNCDI IV. This project has been funded by UEFISCDI under the Driving Urban Transitions Partnership, which has been co-funded by the European Commission.